\documentclass[12pt,a4paper]{article}
\usepackage[T1]{fontenc}
\usepackage[utf8]{inputenc}
\usepackage{authblk}
\usepackage{amsmath,amssymb,epsfig,cite,dsfont}
\usepackage{array,multirow}
\usepackage {color}
\usepackage[toc,page]{appendix}
\numberwithin{equation}{section}
\usepackage{afterpage}
\usepackage{cite}

\definecolor{verde}{cmyk}{.83,.21,1,.08}

\usepackage[colorlinks=true
,urlcolor=blue
,anchorcolor=blue
,citecolor=blue
,filecolor=blue
,linkcolor=blue
,menucolor=blue
,pagecolor=blue
,linktocpage=true
,pdfproducer=medialab
,pdfa=true
]{hyperref}

\advance\hoffset by -1.1truecm
\advance\voffset by -1.0truecm
\newcommand{\be}{\begin{equation}}
\newcommand{\ee}{\end{equation}}
\newcommand{\bea}{\begin{eqnarray}}
\newcommand{\eea}{\end{eqnarray}}

\newcommand{\da}{{\mathcal{D}(\mathcal{A})}}
\newcommand{\gci}{{\mathcal{G}_0^\infty}}
\newcommand{\gc}{{\mathcal{G}_0}}
\newcommand{\mink}{\overline{M}_4}
\newcommand{\cD} {$\mathcal{D}$}
\newcommand{\cF} {$\mathcal{F}$}

\newcommand{\cA} {$\mathcal{A}$}
\usepackage{mathrsfs}
\usepackage{authblk}

\numberwithin{equation}{section}
\usepackage{float}
\restylefloat{figure}
\newcounter{appendice}

\begin{document}

\title{
\vspace{-2.2cm}
\hfill {\small ICCUB-16-042}
\begin{center} \bf{Equations of Motion as Constraints:\\
Superselection Rules, Ward Identities
\linethickness{.05cm}\line(1,0){433}
}\end{center}}
\author[1]{M. Asorey\thanks{asorey@unizar.es}}
\author[2,3]{A.~P.~Balachandran\thanks{balachandran38@gmail.com}}
\author[4,5,6] {F. Lizzi\thanks{fedele.lizzi@na.infn.it}}
\author[4,5]{ G. Marmo,\thanks{marmo@na.infn.it}}
\affil[1]{\small Departamento de F\'{\i}sica Te\'orica, 
Universidad de Zaragoza,  Zaragoza, Spain.}
\affil[2]{\small Physics Department, Syracuse University, Syracuse, New York,  U.S.A.}
\affil[3]{\small Institute of Mathematical Sciences, C.I.T Campus, Chennai, India}
\affil[4]{\small  Dipartimento di Fisica ``E. Pancini'' Universit\`a di
Napoli {\sl Federico II},  Napoli, Italy. }
\affil[5]{\small  INFN Sezione di
Napoli,  Napoli, Italy. }
\affil[6]{\small Departament de Estructura i Constituents de la Mat\`eria, Institut de Ci\'encies del Cosmos, Universitat de Barcelona, Barcelona, Spain}

\renewcommand\Authands{ and }

\date{} 
\maketitle

\begin{abstract}
The meaning of local observables is poorly understood in gauge theories, not to speak of quantum gravity. As a step towards a better understanding we study asymptotic (infrared) transformations in local quantum physics. Our observables are smeared by test functions, at first  vanishing at infinity. In this context we show that the equations of motion can be seen as constraints, which generate a group, the group of space and time dependent  gauge transformations. This is one of the main points of the paper. Infrared nontrivial effects are captured allowing test functions which do not vanish at infinity. These extended operators generate a larger group. The quotient of the two groups generate superselection sectors, which differentiate different infrared sectors.
The BMS group  changes the superselection sector, a result long known for its Lorentz subgroup. It is hence spontaneously broken.
Ward identities implied by the gauge invariance of the S-matrix  generalize the standard results and lead to charge conservation and low energy theorems. Their validity does not require Lorentz invariance.

\end{abstract}

\section{Introduction}

In this paper, we approach the study of asymptotic (infrared) transformations using the concepts of local quantum physics. Since the meaning of local observables is poorly understood in quantum gravity, we focus instead on quantum electrodynamics. Although the theory is well understood, quantisation on spatial slices and its variants seem inadequate for a satisfactory conceptual basis when one explores the limit of low energy photons. A spacetime approach to quantisation is therefore suggested, and we will follow it. Low energy results have  recently attracted attention from investigators of low energy theorems~\cite{Strominger, Strominger1, Campiglia-Laddha,Strominger2, Gomez, Porrati,GomezPanchenko, Strominger3}. 

The focus in this paper is on the covariant formulation of Gauss law and infrared effects. For these reasons we concentrate first on the free Maxwell equations. That bring out the results of interest to us.
Following the work of Peierls in 1952  \cite{Peierls}, we introduce commutators for smeared fields, defining the algebra $\mathcal{A}$ of electromagnetic observables. The Maxwell equations are formulated as quantum constraints. They are defined by operators $G(\eta)$ depending on  test functions $\eta_{\mu}$ vanishing at infinity. They are  the spacetime analogues of the Gauss law operator $\nabla^i E_i$ for the electric field $E_i$ in canonical quantisation, and generate spacetime dependent gauge transformations. They are first class in the sense of Dirac \cite{Hanson-Regge-Teitelboim} and vanish on the domain $\mathcal{D}(\mathcal{A}) $ of $\mathcal{A}$. Following the terminology of the canonical approach, we call the gauge group they generate on exponentiation as  $\mathcal{G}_0^\infty$  \cite{classical_topology_book}, where the superscript indicates that $\eta$ vanishes at infinity and the subscript indicates that the group is connected to the identity.

Infrared effects are captured allowing test functions $\zeta$ which do not vanish at spatial infinity. That leads to operators $Q(\zeta)$, which  are also constructed from the equations of motion. When $\zeta$ is chosen to have the proper infrared behaviour, they become
$G(\zeta)$. The group that $Q(\zeta)$ generate on exponentiation is called   $\mathcal{G}_0$.

The operators $Q(\zeta)$ need not vanish on $\mathcal{D}(\mathcal{A})$ if $\zeta$ does not vanish at infinity\cite{classical_topology_book}. The group $\mathcal{G}_0^\infty$ is normal in $\mathcal{G}_0$. It is the quotient group $\mathcal{G}_0/\mathcal{G}_0^\infty$ that acts on $\mathcal{D}(\mathcal{A})$ effectively, since $\mathcal{G}_0^\infty$ acts as identity. But  it is also the case that $Q(\zeta)$ commutes with elements of $\mathcal{A}$ so that $\mathcal{G}_0\mathcal{D}(\mathcal{A})\subseteq\mathcal{D}(\mathcal{A})$.
The representation of $\gc/\gci$ on $\da$ is an invariant of the representation of \cA\ on $\da$ and defines a superselection sector.

The Lie algebra of $\gc/\gci$ consists of real functions $\widetilde{\rho}$ on $S^2$ and is an abelian group. The $S^2$ emerges from the infrared cutoff, blowing up the tip of the light cone in momentum space:
$$\widetilde{\zeta}(\hat{k})=\lim_{k\to 0} \widetilde{\zeta}(k_0,\, k\,\hat{k}).$$
The group is isomorphic to the Sky group $\gc/\gci$ introduced by one of us (APB) and Vaidya in~\cite{BV}, but in the latter the sphere arises from blowing up spatial infinity.

We then consider gauge invariance, Ward identities and low energy theorems. Since $Q(\zeta)$ and $G(\eta)$ generate gauge transformations and hence commute with all observables, they commute also with the S-matrix $S$. This is the familiar statement that $S$ is gauge invariant. But in contrast to the usual treatments, we have the operator realisation of spacetime gauge transformations. That is important: if these operators do not exist, the proof of the gauge invariance of QED will not be complete.

As is well-known, from the gauge invariance of the $S$-matrix, Ward identities follow. But we can also
deduce low energy theorems therefrom. Thus if $\zeta(\hat{k})=1$, $[Q({\zeta}),S]=0$ gives charge conservation, while other choices of $\zeta(\hat{k})$ lead to other low energy theorems. We do not use Lorentz invariance to deduce charge conservation [cf. \cite{Weinberg-book}] since because of infrared effects, the Lorentz group is spontaneously broken in QED: it cannot be unitarily implemented \cite{Roepstorff,Frohlich,Morchio-and-Strocchi,Buchholz}.

Incidentally, since charge conservation comes from the behaviour of QED as the photon frequency goes to zero, i.e.\ at large distances, it is appropriate to call it as a {\it low energy theorem}.

The BMS group first arose as an asymptotic group in the analysis of asymptotically flat gravity. It was later understood as the group which acts on the null infinities $\mathcal{J}^\pm$ of the 
conformally compactified Minkowski space $\mink$. We argue that it acts as automorphisms on $\gc/\gci$, but not on $\gc$ or $\gci$ separately. Further this action is non-trivial and changes the {\it eigenvalues} of $Q(\zeta)$. In that manner the BMS group action changes the superselection sector. By definition, then it is spontaneously broken. This result, as mentioned above, is known for its Lorentz subgroup, but  is here  also extended to supertranslations. 

Low energy theorems involving photon were already present in the work of Low \cite{Low54} and Gell-Mann and Goldberger \cite{gellgold54}, they are described in~\cite[Chaps.\ 7 and 11]{Nair}. They proved that the zero energy total cross section in Compton effect
is exactly given by the Thompson formula. This work was generalised by Singh \cite{sing67} to obtain subleading terms in the photon frequency. These results are based on analyticity and are non-perturbative, being valid in the Standard Model. and beyond. Later there appeared many low energy theorems from the Goldstone modes of chiral symmetry breaking and applied to pion scattering amplitudes (see\cite{Gundssik,Nicodemi,Weinberg} and references therein). These {\it theorems} in contrast to the above work in photons, were only approximate as they treated the pions as particles of zero mass.

\section{Equations of motion as the Covariant Gauss law}

In this section we will argue that the equations of motion of electrodynamics can be interpreted as constraints. We restrict ourselves to free electromagnetism until Sec.~\ref{GaugeinvWard}.

\subsection{Smearing}

The observables we should consider are the quantum connections \cA, but it is necessary to ``smear'' them with appropriate test functions. We will do this  following Roepstorff~\cite{Roepstorff},  the algebra  of electromagnetic observables is taken to be generated by quantum connections $A_\mu$ smeared with test functions $f^\mu$, which are real, smooth, vanishing at  infinity and   obey the Lorentz gauge condition $\partial_\mu f^\mu=0$, i.e. 
\begin{equation}
A(f)=\int d^4 x\, f^\mu(x) A_\mu(x);\quad f^\mu \in C_0^\infty({\mathbb R}^4);\  \partial_\mu f^\mu=0.
\label{test}
\end{equation}
Notice that the elements  \cA\  are gauge invariant because of the transversal character of the test functions. 

Usually test functions are taken to be of compact support. For the purposes of this paper this requirement is too drastic. We certainly need the functions to vanish at infinity, but the useful requirement is in reality on the behaviour of the Fourier transform for small momenta. We define, for $k^\mu k_\mu=0$ (on the mass shell):
\begin{equation}
\label{five}
\widetilde{f}^\mu ({k})=\int d^4 x\,{f}^\mu ({x})\, e^{-i k\cdot x}.
\end{equation}
Let us introduce the space $\mathcal C$ of functions which satisfy the following constraints:
\begin{equation}
\tilde f_\mu(\mathbf{k}),\ \partial_{i_1}\cdots \partial_{i_n}\,\partial^\mu \tilde f_\mu (\mathbf{k}),\  n=1,2,3, \dots
\label{ttt}
\end{equation}
are all finite as $\mathbf{k}\to 0$. Hence in particular
\begin{equation}
\lim_{\mathbf{k}\to 0}k\cdot\tilde f(\mathbf{k})=0 
\label{t4}
\end{equation}
\begin{equation}
\lim_{\mathbf{k}\to 0} \partial_{i_1}\cdots \partial_{i_n} \ k\cdot \tilde f (\mathbf{k})< \infty,\ \  \mathrm{for}\ n=1,2,3,\dots
\label{t5}
\end{equation}

Compact support functions belong to $\mathcal C$, but the space contains also functions which do not vanish fast for $x\to\infty$, and therefore are not of compact support. We require $f_\mu\in \mathcal C$.  For us the space $\mathcal C$ will play the role usually played by compact support functions. 

Upon quantization, $A_\mu$ has the mode expansion
\begin{equation}
\label{modes}
A_\mu(x)=\int d\mu(\mathbf{k})[a_\mu (\mathbf{k}) e^{-i k\cdot x} + a_\mu (\mathbf{k})^\dagger e^{i k\cdot x}].
\end{equation}
with $k_0=\sqrt{\mathbf{k}^2}=|\mathbf k|$ and the usual invariant measure is 
\begin{equation}
 d\mu(\mathbf{k})= \frac{d^3 \mathbf{k}}{(2\pi)^3 2 k_0} \ . \label{measure}
\end{equation}
With standard  commutation relations for $a_\mu$ and $a_\mu^\dagger$, $A(f)$ acts on the Fock space  \cF\ defined as
the Hilbert space completion of the multiphoton space of states
\begin{equation}
\mathcal{F}=\overline{\bigoplus_{n=0}^\infty\ ^s\mathcal{H}^{{\otimes   n}}}.
\label{Fock2}
\end{equation} 
where the $\mathcal H^{\otimes n}$ are generated by the action of $n$ creation operators  $a_\mu (\mathbf{k})^\dagger$ and the ${}^s$ indicates symmetrized states.
To see the action of  $A(f)$, one notes that it depends on 
\begin{equation}
\label{Fourier}
\int d\mu(\mathbf{k})\, \widetilde{f}^\mu  ({k}) a_\mu  (\mathbf{k}),
\end{equation}
and its adjoint 
\begin{equation}
\int d\mu(\mathbf{k})\, \widetilde{f}^{\mu}  ({k})^\ast a_\mu^\dagger  (\mathbf{k}).
\end{equation}

The commutator $[A(f),A(g)]$ for two such test functions $f$ and $g$ is
\begin{equation}
[A(f),A(g)]=\int\! d^4 x\! \int\! d^4 y\,f^\mu(x) D(x-y) g_\mu(y),
\end{equation}
where $D$ is the causal  Pauli-Jordan function
\begin{equation}
 D(x-y) = \int d\mu(\mathbf{k}) [e^{-i k\cdot(x-y)}-e^{i k\cdot(x-y)}]
\end{equation}
The causal  function $D$ satisfies the wave equation
\begin{equation}
\label{boxD}
 \square\, D(x)=0.
\end{equation}

The unitary operators $W(f)=e^{i A(f)}$ generate a Weyl algebra $\mathcal{W}$. From the mathematical point of view, it would be better to work with $\mathcal{W}$. But we choose to work with $A(f)$ which is better known in physics. Notice that the domain of the algebra $\mathcal{A}$ in the Fock space $\mathcal{F}$ representation is smaller that that of  $\mathcal{W}$ which is the the full Fock space.

We next consider the equations of motion. Classically they are 
\begin{equation}
\partial^\lambda F_{\lambda\mu}(A)=0,\quad \hbox{with}\  F_{\lambda\mu}(A)=\partial_\lambda A_\mu
-\partial_\mu A_\lambda.
\end{equation}
We must smear the corresponding operator with test functions 
$\eta_\mu\in C^\infty_0(\mathbb{R}^4)$ and transfer derivatives to $\eta$ to get a sensible quantum operator, or even a sensible generator of canonical transformations. Otherwise distributions like  $D$, or worse, will occur in commutators or Poisson brackets.

Towards this end, let us consider  
$F_{\mu\nu}(\eta)=\partial_\mu\eta_\nu-\partial_\nu\eta_\mu$ in (\ref{test})  for a test function $\eta_\mu\in \mathcal C$, {and the smearing of the equation of motion},
\begin{equation}
\label{gauss}
G(\eta)=\int d^4x\, \partial^\lambda F_{\lambda\mu}(\eta) A^\mu, \quad \eta_\mu\in C^\infty_0(\mathbb{R}^4) .
\end{equation}
The test function for $A_\mu$ that appears   in (\ref{gauss}) is not $\eta_\mu$, but $\partial^\lambda F_{\lambda\mu}(\eta)$ that fullfils (\ref{test}) and the Fourier transform condition~\eqref{ttt}.

The following properties of $G(\eta)$ may be noted:
\begin{itemize}

\item[a)] For classical fields $A_\mu$, partial integration gives 

\begin{equation}
\label{gauss2}
G(\eta)=\int d^4x\, \eta^\mu\partial^\lambda F_{\lambda\mu}(A),
\end{equation}
which is zero by equations of motion. Hence in quantum physics, we can set it as a constraint on the domain $\mathcal D(\mathcal A)$ of \cA.
\begin{equation}
\label{gauss3}
G(\eta)|\psi\rangle=0 \qquad \mathrm{if}\ \   |\psi\rangle\ \in\ \hbox{\cD (\cA)}.
\end{equation}

\item[b)] 
Consistency demands that $G(\eta)$ are first class constraints. That is the case, for we have 
 \begin{equation}
\label{gauss4}
[G(\eta_1),G(\eta_2)]= \int d^4x\, d^4y \, \partial^\lambda F_{\lambda\mu}(\eta_1)(x) D(x-y) \partial^\rho F_\rho^{\mu}(\eta_2)(y)= 0,
\end{equation}
for any pair $\eta_{1,\mu},\eta_{2,\mu} \in C^\infty_0(\mathbb{R}^4)$.
Here 
 \begin{equation}
\label{eom}
 \partial^\lambda F_{\lambda\mu}(\eta_a)=\square \,\eta_{a,\mu}-\partial_\mu (\partial^\lambda\, \eta_{a,\lambda})
 \qquad a=1,2.
\end{equation}
The $\square$ term vanishes after partial integration and use of $\square D=0$ as in (\ref{boxD}).
The $\partial_\mu$ term vanishes after partial integration and use of $\partial_\mu D(x-y)=-D(x-y)\partial_\mu$, $F_{\rho \mu}(\eta_2)$ being anti-symmetric.

\item[c)] 
Consistency also demands that \cA\cD(\cA)$\subseteq$ \cD(\cA)\  or that \cA\ are {\it first class variables}. That is also the case. We show that in two steps. The first demonstrates that $G(\eta)$ generates gauge transformations. From there follows the second result, $[G(\eta),A(f)]=0$.
Indeed
 \begin{equation}
\label{gauss5}
[G(\eta),A_\mu(x)]\!=\!  \int d^4y \, \partial^\lambda F_{\lambda\mu}(\eta)(y) D(y-x)\!=\!
-\partial_\mu \int d^4y \, (\partial^\lambda \eta_\lambda)(y) D(y-x)\!:=\! i\partial_\mu \Lambda(x).
\end{equation}
because the box term again  vanishes by (\ref{boxD}). So $G(\eta)$ is the generator of a gauge transformation.
\end{itemize}
{This establishes the connection between the equations of motion and the constraints. We see that the role of the smearing functions, and of their infrared behaviour, is fundamental.}

 We can now verify (\ref{gauss3}) trivially in \cD(\cA) in the Fock space representation.
  Use (\ref{Fourier}) and (\ref{five}) to write 
  \begin{equation}
\label{Fourier2}
A_\mu(x)=  \int d\mu(\mathbf{k})[a_\mu (\mathbf{k}) e^{-i k\cdot x} + a_\mu (\mathbf{k})^\dagger e^{i k\cdot x}],\quad k^2=0,\ k^0>0,
\end{equation}
with the constraint 
\begin{equation}
 k^\mu a_\mu(\mathbf{k})|\psi\rangle\ =0,
 \label{constrain}
 \end{equation}
which we can assume by the gauge invariance of (2.1) and  (2.16).Thus, for any Fock state $|\psi \rangle\in$ \cF, we find that 
 \begin{align}
\label{Gausspi}
G(\eta)|\psi\rangle=& \nonumber  \int d^4 x \, \partial^\lambda F_{\lambda\mu}(\eta)(x) A^\mu(x) |\psi\rangle\ \\ =&  
\!-\! \int d\mu(\mathbf{k})\left[ [k^2 \widetilde{\eta}_\mu(\mathbf{k})-k_\mu\  k\cdot  \widetilde{\eta}(\mathbf{k})]a^\mu(\mathbf{k})+[k^2 \widetilde{\eta}^\mu(\mathbf{k})^\ast-k^\mu\ k\cdot  \widetilde{\eta}(\mathbf{k})^\ast]a^\dagger_\mu(\mathbf{k})\right] |\psi\rangle\ \nonumber\\=&0.
\end{align}
where
\begin{equation}
\label{six}
\widetilde{\eta}_\mu ({\mathbf k})=\int d^4 x\,{\eta}_\mu ({x})\, e^{-i k\cdot x},
\end{equation}
and $k^0=\sqrt{\mathbf k^2}$.
The group generated by $G(\eta)$ is denoted by $\gci$. It acts trivially in \cF.

\noindent \textit{Remark:} {With  hindsight we have here just verified that the Fock space of physical states \cF\ is in the kernel of $G(\eta)$. This is crucial for the interpretation of the equations of motion as constraints.}

\smallskip

One characteristic  operator of Fock space  \cF\ is  the number operator
\begin{equation}
\label{seven}
N=\int d\mu(\mathbf{k}) N(\mathbf{k}),
\end{equation}
where 
\begin{equation}
\label{seventh}
N(\mathbf{k})=a_\mu(\mathbf{k})^\dagger a^\mu(\mathbf{k}).
\end{equation}
The domain of $N$ is given by the states $|\psi\rangle\in$\cF\
such that 
$$ \langle \psi|N^2|\psi\rangle< \infty.$$
This domain is a subset of the larger set of states where $N$ has finite
expectation values, i.e. $ \langle \psi|N|\psi\rangle< \infty.$
Because of the constraint (\ref{constrain}) $N$ is a positive operator in \cF. 

\subsection{Vacua and Coherent States}

Let us now consider the coherent state defined by
 \begin{equation}
\label{eight}
|f\rangle=e^{iA(f)}|0\rangle,
\end{equation}
with $f\in\mathcal C$ and
\begin{equation}
\label{nine}
(f,f):= \int d\mu(\mathbf{k}) \widetilde{f}^\mu(\mathbf{k})^\ast\widetilde{f}_\mu(\mathbf{k})<\infty,
\end{equation}
i.e. $|f\rangle\in\,$\cF . It is easy to show that 
$(f,f)\geq 0$. This is a consequence of the transversal and small $k$ properties of $f$.
Indeed, since $\partial^\mu f_\mu=0$,  $k^\mu \widetilde{f}_\mu=0$, and since $k_0=\sqrt{\mathbf{k}^2}$
it follows that $\widetilde{f}_\mu^\ast \widetilde{f}^\mu>0$ unless ${f}_\mu=0$ or 
$\widetilde{f}_\mu(\mathbf{k})=k_\mu \widetilde{\phi}(\mathbf{k})$. In the latter case $f_\mu=\partial_\mu
\phi$, but by the transversality condition $\square{\phi}=0$, which since $\phi$ is compactly supported implies that $\phi=0$ and, thus,  ${f}_\mu=0$.

Also
  \begin{eqnarray}
\label{vev}
\langle f| A_\mu (x)|f\rangle &=& -i  \nonumber  \int d^4 y \,  f_\mu(y) D(y-x) \\ &=& \!\!\! 
\nonumber
-i \int d\mu(\mathbf{k})\ [\widetilde{f}_\mu ({k})\, e^{i k\cdot x}- \widetilde{f}_\mu ({k})^\ast\, e^{-i k\cdot x}].
\end{eqnarray}
 Thus from the mode expansion of $A_\mu$ in (\ref{modes}), we see that
   \begin{equation}
\label{vev2}
\langle f| a_\mu (\mathbf{k})|f\rangle =i{\widetilde{f}_\mu}(\mathbf{k})^\ast .
\end{equation}
 Hence
   \begin{equation}
\label{vev3}
0\leq\langle f| N |f\rangle =(f,f)< \infty .
\end{equation}
 We now remove the requirement that $f\in\mathcal C$ and vanishes at infinity in $x$,  replacing $e^{iA(f)}$ by $e^{iA(g)}$ in (\ref{eight}) 
 where $g$ is a transverse, but not vanishing at infinity, function of $C^\infty(\mathbb{R}^4)$ with
\be
\lim_{\varepsilon\to 0} \int_{k_0 >\varepsilon}d\mu(\mathbf k) \bar g(k_0,\mathbf k) g(k_0,\mathbf  k) =\infty\ .  
\ee
We assume that $g(k_0,{\mathbf k})$ is $O(1/k_0)$ for $k_0$ going to 0, so that the divergence is at worst logarithmic. Then the expectation value of  $N$  in the state $|g\rangle$ diverges: it has an infinite number of infrared photons,
    \begin{equation}
\label{vev4}
\langle g| N |g\rangle = \lim_{\epsilon\to 0} \langle g| \int_{k_0\geq \epsilon} d\mu(\mathbf{k}) N(\mathbf{k})  |g\rangle=\infty.
\end{equation}
while its energy 
\begin{equation}
\label{energy}
E=\lim_{\epsilon\to 0}  \int_{k_0\geq \epsilon} d\mu(\mathbf{k}) k_0  \bar g(k_0,\mathbf k) g(k_0,\mathbf k) 
\end{equation}
can remain finite.

The state $|g\rangle$ built by  applying $\exp(i A(g)$ on the vacuum,   generates a state which does not belong to  the domain of the number operator $N$. 
As is known \cite{Morchio-and-Strocchi,Eriksson,Kibble,Kibble2, Kibble3,Kibble4} and we shall later see in Section 5, infrared dressing does not leave invariant the domain of the number operator in  Fock space.

\section{The Superselection Algebra}

Suppose next that we replace the test functions $\eta_\mu$ in  $G(\eta)$ 
by $\zeta_\mu$ which need not vanish at infinity, and therefore does not belong to $\mathcal C$. This defines a new class of operators, analogs of the $G$'s, which we will indicate with a different symbol, to stress the difference: 
\begin{equation}
Q(\zeta)=\int d^4 x\ \partial^\lambda F_{\lambda \mu} (\zeta) (x) A^\mu(x)\ .
\end{equation}
They reduce to $G(\eta)$ if $\zeta_\mu=\eta_\mu$ belongs to $\mathcal C$: $Q(\eta)=G(\eta)$. But in general $Q(\zeta)\neq G(\zeta)$ if $\zeta$ is not in $\mathcal C$, and, thus, $Q(\zeta)$ need not vanish on \cD(\cA).

The operators $Q(\zeta)$ commute with both $G(\eta)$  and $A(f)$ as before since both $\eta$ and $f$ belong to $\mathcal C$, letting us do the needed partial integrations:
\begin{equation}
[Q(\zeta),G(\eta)]= [Q(\zeta),A(f)]=0\ .
\end{equation}
So the irreducible representations of $A(f)$ where $G(\eta)$  vanishes can be labelled by the irreducible representations of $Q(\zeta)$: the latter are superselected.

The  $Q(\zeta)$  generates on  exponentiation the group $\gc$. Its subgroup $\gci$ is normal in $\gc$, commuting with all elements of $\gc$. Further $\gci$ acts as identity on \cD(\cA). Hence the group classifying superselection sectors is $\gc/\gci$. 

We now comment briefly on the asymptotics of $\eta_\mu$ and $\zeta_\mu$ .

\subsubsection*{Remarks on asymptotics}

The large $x$ behaviour of $\eta_\mu$ and $\zeta_\mu$ controls the behaviour of $\widetilde{\eta}_\mu(\mathbf{k})$ and $\widetilde{\zeta}_\mu(\mathbf{k})$
as $k_0\to0$. We now explain this point.

Since $\zeta$ is not of compact support the behaviour of its Fourier transform $\tilde\zeta(\mathbf k)$ near the origin is different,  it may diverge at the origin and
 we shall later see that~(\ref{t4}) can be replaced by 
\begin{equation}
k\cdot\widetilde{\zeta}(\mathbf{k})
 \xrightarrow[\mathbf{k}\to 0]{} \frac{\widetilde{\zeta}(\widehat{\mathbf k})}{|\mathbf k|^{\alpha}}\  \mathrm{with}\ \alpha\leq 2,
\label{t6}
\end{equation}
where $\widehat{\mathbf k}={\mathbf k}/{ |\mathbf k|}$.
This conclusion is reached by requiring that $Q(\zeta)$ acts without divergent terms on the infrared dressed charged states.
When 
\begin{equation}
\lim_{\mathbf{k}\to 0}k\cdot\widetilde{\zeta}(\mathbf{k})\neq 0, %
\label{t55}
\end{equation}
 $\zeta_\mu$ cannot have compact support as shown by the derivation of (\ref{t4}) (\ref{t5}).

Since $\gci$ has elements with the properties (\ref{t4}), (\ref{t5}), we can say that 
\begin{equation}
\nonumber
Q(\zeta^{(1)})-Q(\zeta^{(2)})= G(\zeta^{(1)}-\zeta^{(2)})
\label{t7}
\end{equation}
if
{\begin{equation}
\nonumber
\widetilde{\zeta}^{(1)}-\widetilde{\zeta}^{(2)}\in\mathcal C
\label{t8}
\end{equation}
}

Since 
\begin{equation}
G(\zeta^{(1)}-\zeta^{(2)})\hbox{\cD(\cA)}=\{0\}.
\label{t9}
\end{equation}
we can identify all such $\zeta$'s differing  by an $\eta$ as $\mathbf{k}\to 0$.
The conclusion is that in view of  (\ref{t6}) elements of $\gc/\gci$ are characterised  by functions on $S^2$ and the index $\alpha$.

For fixed $\alpha$. the group $\gc/\gci$  is thus isomorphic to the gauge group of maps from $S^2$ to $U(1)$.
\begin{equation}
\gc/\gci=\hbox{Maps($S^2,{U(1)}$)}.
\label{t99}
\end{equation}

A natural question to ask is whether  $\gc/\gci$ is abelian or not. 
The answer  depends on the commutator $[Q(\zeta_1), Q(\zeta_2)]$.
 We find from (\ref{vev}) after changing $\eta_1$ to $\zeta_1$
 \begin{equation}
[Q(\zeta_1), Q(\zeta_2)]=\!\! \int\! d\mu(\mathbf{k}) \left[ \left(k^\mu k\cdot \widetilde{\zeta_1}(\mathbf{k})\right)\!\left(k_\mu k\cdot \widetilde{\zeta}_2(\mathbf{-k})\right)\!+\left(k^\mu k\cdot \widetilde{\zeta_1^\ast}(\mathbf{k})\right)\!\left(k_\mu k\cdot \widetilde{\zeta}^\ast_2(\mathbf{-k})\right)\right].
\label{t12}
\end{equation}
For $\alpha<1$ the coefficient integral of $k\cdot k$ converges in the infrared because  the measure is
$d\mu(\mathbf k) = d |\mathbf k| d\theta d \phi |\mathbf k| \sin\theta/(2\pi)^3$. As $k \cdot k=0$, $[Q(\zeta_1), Q(\zeta_2)]=0$ if both $\zeta_1$ and $\zeta_2$ have the same $\alpha<1$. The integral is also zero if $\widetilde{\zeta}$ is odd in $\mathbf{k}$. Otherwise, or if $\alpha>1$, the integral diverges.
 The divergence for $\alpha=1$ of the above integral is logarithmic. There may be a regularisation to get a finite answer even if $\widetilde{\zeta}$ is constant.
 We can treat pairs $\widetilde{\zeta_1}$,$\widetilde{\zeta_2}$ with different $\alpha$. in the same manner.
 In the divergent cases $Q$'s do not form a Lie algebra. Such domain problem may not spoil physics.

\section{Gauge Invariance and Ward identities\label{GaugeinvWard}}

A concise statement of gauge invariance and Ward identities as formulated in textbooks is the following: let $S_I$ be the interaction representation $S$-matrix in QED:
\begin{equation}
\label{ward}
S_I=T\exp i\int d^4x\, A_{\mu}J^\mu.
\end{equation}
Here $J_\mu$ is a conserved current\footnote{We are now going beyond free QED to a slightly more general case.}.
Then $S_I$ is invariant under the gauge transformation
\begin{equation}
\label{gauge}
A_\mu\to A_\mu +\partial_\mu \Lambda
\end{equation}
so that
\begin{equation}
\label{inv}
S_I=T S_I \exp i\int d^4x\, J^\mu\partial_{\mu} \Lambda.
\end{equation}
The matrix elements of (\ref{inv}) between the initial and final states are also gauge invariant. 

Expanding (\ref{inv}) in powers of the coupling constant and taking its matrix elements, one gets Ward identities order by order.

This treatment is not satisfactory {for our purposes}. The first point  is that the operator implementing (\ref{gauge}) in the whole  {\it spacetime} (and not just as at constant time, as in the $A_0=0$ gauge) is not shown.

The more serious problem is that the initial and final states are {considered to have} a finite number of photons. But because of infrared effects, it is known that this is not to be the case. Below we focus on just this infrared part of $S_I$, and outline the approach of Roepstorff~\cite{Roepstorff}, which is supported by a considerable literature (c.f.~\cite{Eriksson},~\cite{Frohlich} and references therein).

\subsubsection*{Infrared dressing of initial state} 

Consider the  initial state with one particle\footnote{A simple modification of what follows also covers the case of several charged particles. The results depend only on the total momentum and total charge of the multiparticle system.
} and no photons,
\begin{equation}
\label{Fock1}
|0\rangle_\gamma | p,e\rangle,
\end{equation}
where $|0\rangle_\gamma$ is the photon Fock vacuum and $| p,e\rangle$ the state of a charged particle of momentum $p$ and charge $e$. One has to take into account the fact that charged particles radiate. If $H_I$ is the interaction Hamiltonian, the {\it in} state is 
\begin{equation}
\label{in}
T\exp \left\{ i\int_{-\infty}^0 dx_0 H_I \right\} |0\rangle_\gamma | p,e\rangle,
\end{equation}
so that this radiation has accumulated for an infinite period of time. We want to approximate the effect of this $H_I$ due to vanishingly small photon frequency $k_0$.

\subsubsection*{The infrared model}

The current $J^{\mu}$ for the infrared model is that of a charged particle of charge $e$, mass $M$ and constant momentum  $p^\mu$. Thus in the current
\begin{equation}
\label{model}
J^{\mu}(\mathbf{k})= e\int d\tau\, \dot{\zeta^\mu}\, \delta^4(x-\zeta(\tau)),
\end{equation}
of classical electromagnetism, we set
\begin{equation}
\label{notation}
{\zeta^\mu}(\tau)=\frac{p^\mu}{M} \tau
\end{equation}
Thus the change of particle momentum due to the back reaction to photon emission is neglected in the approximation of interest of large $M/k_0\gg 1$.
Substituting  (\ref{notation}) in (\ref{model}), we get the current 
\begin{equation}
\label{model2}
J^{\mu}(\mathbf{k})= e\int d\tau\, \frac{p^\mu}{M} \, \delta^4\left(x-\frac{p}{M}\tau\right),
\end{equation}
and the interaction Hamiltonian
\begin{equation}
\label{ward2}
H_I(x_0)=\exp \int d^3x\,J^\mu(x)  A_{\mu}(x).
\end{equation}
The infrared {\it in} state is then easily calculated \cite{BV}:
\begin{equation}
\label{in2}
 | p,e;\gamma \rangle_\mathrm { in}=T \exp \left\{i\int_{-\infty}^0 dx_0 H_I(x_0) \right\} |0\rangle_\gamma | p,e\rangle,
\end{equation}
Here the time ordering is not needed in the right hand side, since the commutator 
$[H_I(x_0),H_I(x_0')]$ is a multiple of identity.

We can write (\ref{in2}) in an elegant from
\begin{equation}
\label{in3}
 | p,e;\gamma \rangle_\mathrm { in}= \exp \left\{i e\int_{-\infty}^0 d\tau\,  \frac{\ p^\mu}{M}\, A_\mu\left({\tau p}/{M}\right) \right\}\,  |0\rangle_\gamma | p,e\rangle,
\end{equation}

The exponential is just the Wilson line integral along the particle trajectory. A generalization of the same approximation for non-abelian gauge theories gives some hints on the quark confinement mechanism \cite{as,as2}.

\subsubsection*{Gauge properties of infrared dressed state}
By \emph{gauge invariance} is meant invariance of~(\ref{model2}) under $G(\eta)$, while the response
under $Q(\zeta)$ defines its superselection sectors. 
Therefore we have to show that 
\begin{equation}
\label{in4}
e^{iG(\eta)} | p,e;\gamma \rangle_\mathrm { in}=  | p,e;\gamma \rangle_\mathrm {in}.
\end{equation}
Now 
\begin{eqnarray}
\label{in5}
\!\!e^{iG(\eta)} | p,e;\gamma \rangle_\mathrm { in} =  
\left[e^{iG(\eta)} \exp \left\{\!-i\!\! \int_{-\infty}^0\!\!\! dx_0\!\!\int d^3 x J^\mu(x) A_\mu(x)\right\} e^{-i G(\eta)}\right]  e^{i G(\eta)}   |0\rangle_\gamma | p,e\rangle
\end{eqnarray}
The factor within the brackets is  
\be
\label{in6}
  \exp\left\{-i \int_{-\infty}^0 dx_0 \int d^3 x J^\mu(x) \partial_\mu\Lambda(x)\right\}
\ee
which  becomes after integration by parts
\begin{eqnarray}
\label{in66}
 && \exp \left\{-i \int_{x_0=0} d^3 x J^0(x) \Lambda(x)\right\}=e^{-i e\Lambda(0)},
\end{eqnarray}
 since  ({\ref{model2}})
\begin{equation}
\label{in7}
 J^0(0, {\mathbf{x}})= e\ \delta^3({\mathbf{x}}) %
 \end{equation}
and  $\Lambda \to 0$ as $x_0\to \infty$.

Thus since the charged particle is at spatial origin at time $0$, the gauge transform of $ | p,e;\gamma \rangle_\mathrm { in}$ is
\begin{equation}
\label{in8}
e^{-i e\Lambda(0)} e^{i G(\eta)}   |0\rangle_\gamma | p,e\rangle\, .
 \end{equation}
The first factor in  (\ref{in8}) comes from gauge transforming $A_\mu$. But for a charge particle at origin, the Gauss law has the additional  term proportional to $J_0=e\, \delta^3({\mathbf{x}}) $, as in $(\partial^i E_i+e\delta^3(\mathbf{x}))|\cdot \rangle=0$. This is to be smeared with $\Lambda (0, {\mathbf{x}})$
to get its contibution to $G(\eta)$. Thus the state $| p,e;\gamma \rangle_\mathrm { in}$ is fully gauge invariant.

\section{The superselection operator $Q(\zeta)$ and charge conservation}

The superselection rules are associated with very large distances and very low frequencies. We can thus choose $\zeta$ to vanish at $x_0=0$ so that $Q(\zeta)$ transforms only the Dirac-Wilson line.

{Before discussing charge conjugation and Ward identities let us discuss the behaviour of the nonvanishing $k\cdot \widetilde{\zeta}( {\mathbf{k}})$ as $k_0\to 0$. 
That implies that  the test function $\zeta\notin\mathcal C$.}

{Let us isolate the angular part and consider the infrared limit 
\begin{equation}
\label{roes2}
k\cdot \widetilde{\zeta}(k)  \xrightarrow[k_0\to 0]{} \frac{\widetilde{\zeta}( {\mathbf{\hat k}})}{k_0^{\alpha}}\qquad \widetilde{\zeta}( {\mathbf{\hat k}})\neq 0.
 \end{equation}
Different values of $\alpha$  show different behaviours, with $\alpha=2$ a separating value, the most interesting case. }

{Expressing the invariant measure~\eqref{measure} as
\begin{equation}
d\mu(\mathbf{k}) =\frac{k_0 d k^0 d \Omega_{\hat{k}}}{(2\pi)^3},
\end{equation}
we can see that the integral (\ref{gauss5}) expressing $\partial_\mu\Lambda$: 
\begin{eqnarray}
i\partial_\mu \Lambda=[Q(\zeta), A_\mu(x)] &=& -\partial_\mu \int d^4 y (\partial^\lambda \zeta_\lambda) (y) D(y-x)\nonumber\\ &=&
-\int d\mu(\mathbf{k}) \left[ k^\mu k\cdot \widetilde{\zeta}({k})e^{ik\cdot x}+ k^\mu k\cdot \widetilde{\zeta}({-k})e^{-ik\cdot x}\right)
\label{comm}
\end{eqnarray}
exists if $\alpha<2$. For $\alpha=2$ the integral exists and is well defined, but the would be function $\Lambda$, obtained by removing $k_\mu$ from 
(\ref{comm}) diverges. The gauge transformation is therefore given by a closed, but not exact form. It would be interesting to study the cohomolgy of this limiting case, but this is beyond the scope of this paper.}

{Likewise the infrared dressed {\it in} state (\ref{in3}) is a state with non-trivial response to $Q({\zeta})$, which depends on the value of $\alpha$. This means these dressed states have values different from zero for the superselection operators~$Q(\zeta)$. We can calculate 
\begin{equation}
e^{i Q(\zeta)} | p,e;\gamma \rangle_\mathrm { in}
\end{equation}
as in (\ref{in4}), noting that 
\begin{equation}
e^{i Q(\zeta)} |0\rangle_\gamma | p,e \rangle=|0\rangle_\gamma | p,e \rangle
\end{equation}
we get 
\begin{eqnarray}
e^{i Q(\zeta)}  | p,e;\gamma \rangle_\mathrm{ in}&=& \exp \left\{ -i e \int^0_{\infty} dx_0\int d^3 x J^\mu(x)   \partial_\mu \Lambda(x)\right\} | p,e;\gamma \rangle_\mathrm{ in}\\
&=&  \exp \left\{ i e \int\! d\mu(\mathbf{k})  \left( k\cdot \widetilde{\zeta}({k})-  k\cdot \widetilde{\zeta}({-k})\right)\right\}| p,e;\gamma \rangle_\mathrm{ in}.
\label{roeb3}
\end{eqnarray}
As $\widetilde{\zeta}$ vanishes fast as $k_0\to\infty$, we have to examine only the $k_0\to 0$ limit.
Thus, the one-particle dressed  states have non-zero values for the superselection operators $Q(\widetilde{\zeta})$. Again  the case $\alpha<2$ poses no problems, as in this case $Q(\zeta) $ is finite. Instead for $\alpha=2$ the exponent of 
(\ref{roeb3}) diverges, unless  $\widetilde{\zeta}$  is odd in $\mathbf{k}$. In this case $| p,e;\gamma \rangle_\mathrm{ in}$ is not in the domain of $ Q(\zeta)$.} 

{We can conclude that the superselection sectors are labelled by $\alpha$, (with $\alpha\leq 2$) and the functions  $\widetilde{\zeta}({\hat{\mathbf k}})$ on $S^2$.}

Some remarks are in order.

The photon momentum $k_0$ for $k_0>0$ lies on a light cone $V_+$ with the tip $k_0=0$ removed. The infrared features we have encountered are all concerned with the limit $k_0\to 0$.
If the tip $k_0=0$ of the light cone  is regarded as just a point, and we denote by $\overline{V}_+$ the light cone  with the tip, any smooth function $\alpha$ on $\overline{V}_+$ will have a constant limit $\alpha(0)$ as
$k_0\to 0$.
But in our case $k_0^{\alpha}  k\cdot \widetilde{\zeta}({k})$ need not be a constant as $k_0\to 0$ since it approaches $\widetilde{\zeta}(0,\mathbf{k})$. This direction-dependent limit can be accommodated by attaching a sphere $S^2$ to  $\overline{V}_+$ and not a point. We have {\it blown\ up} the point to a sphere. This procedure is common in mathematics.

In an earlier work \cite{BV} where the Sky group was introduced, spatial infinity was blown up to a sphere to define this group. Here instead, we get the dual blow-up of the origin in momentum space. For fixed $\alpha < 1$ both groups are isomorphic.  
The Sky group for QED is abelian being the group of maps from $S^2$ to $U(1)$ with natural multiplication
\begin{equation}
e^{i \alpha_i} \in {\rm Sky}; \  e^{i \alpha_1} e^{i \alpha_2}:=e^{i (\alpha_1+ \alpha_2)}
\end{equation}
But $\alpha\geq 1$ requires more discussion (See the discussion of Eq. (\ref{t12})).

 \section{Charge conservation and low energy theorems}

Charge conservation follows directly from the infrared  part in the action of  $e^{i Q(\zeta)} $
on $ | p,e;\gamma \rangle_\mathrm{ in}$. We have~\cite{BalQureshi}
\begin{eqnarray}
e^{i Q(\zeta)}  | p,e;\gamma \rangle_\mathrm{ in}&=& \exp \left\{ -i e \int_{-\infty}^0 d\tau \frac{p^\mu}{M} A_\mu ({\tau p }/{M} )\right\}\nonumber\\
&\times&  \exp \left\{ -i e \int_{-\infty}^0 d\tau  \frac{p^\mu}{M}\partial_\mu \Lambda({\tau p }/{M} ) \right\}| p,e;\gamma \rangle_\mathrm{ }.
\label{roeb5}
\end{eqnarray}
Thus the eigenvalue of $e^{i Q(\zeta)}$ on $| p,e;\gamma \rangle_\mathrm{ in}$ is 
\begin{eqnarray}
\exp&&\!\!\!\!\!\!\!\!\! \left\{ -i e\!\! \int_{-\infty}^0\!\!\!\! d\tau \int\! d\mu(\mathbf{k}) d^4 y\, \frac{p\cdot k}{M} \left[ \partial^\lambda \zeta_\lambda(y)\, e^{-ik\cdot (y-\tau p /M)}+\partial^\lambda \zeta_\lambda(y)^\ast\, e^{ik\cdot (y- \tau p/M)}\right]\right\}=\nonumber
\\
\exp&&\!\!\!\!\!\!\!\!\! \left\{ -i e\! \int\! d\mu(\mathbf{k}) \lim_{\epsilon \to 0} \left[\frac{p\cdot k}{ p\cdot k-i \epsilon} k\cdot \widetilde{\zeta}(\mathbf{k})-\frac{p\cdot k}{ p\cdot k+i \epsilon} k\cdot \widetilde{\zeta}(\mathbf{k})^\ast\right]\right\} =\nonumber \\
\exp&&\!\!\!\!\!\!\!\!\! \left\{ -i e\! \int\! d\mu(\mathbf{k})\left[k\cdot \widetilde{\zeta}(\mathbf{k})-k\cdot \widetilde{\zeta}^(\mathbf{k})^\ast\right] \right\}.
\label{roeb6}
\end{eqnarray}
The integral multiplying $e$ is independent of $p$. Thus for $N$ charged particles of charge $e_i$, $e$ gets replaced by the total charge $q=\sum_{i=1}^N e_i$.
But ${i Q(\zeta)} $ is superselected. Hence its value in the {\it in} state and {\it out} state are the same, so that charge is conserved.

Unlike traditional treatments like~ \cite{Weinberg}, our treatment does not invoke the fact that $A_\mu$ is not a true vector, nor any reference to Lorentz invariance.
The diagrams summed in the infrared dressing operator in  (\ref{in3}) are the sum  of over all photon numbers of the diagrams Weinberg considers. It is the tree approximation to the Feynman diagrams with fixed charge as photon momenta go to zero.

We now remark on going beyond the tree approximation, which is also necessary to get amplitudes which can measure $\partial_\mu \Lambda$. Its presence in (\ref{roeb6}) is such that $\widetilde{\zeta}_\mu$-dependence factors out.

The gauge transformation (\ref{gauss}) (with $\eta\to\zeta$) shifts $J^\mu A_\mu$, which for electron field $\psi$ is the shift of $i\overline{\psi}\gamma^\mu \psi A_\mu$ to $i\overline{\psi}\gamma^\mu \psi \partial_\mu\Lambda$.
This gives a shift of photon creation and annihilation operators

\begin{eqnarray}
a_\mu^\dagger(\mathbf{k})&\rightarrow&  a_\mu^\dagger(\mathbf{k})-i k_\mu k\cdot \widetilde{\zeta}(-k),\nonumber\\
a_\mu(\mathbf{k})&\rightarrow&  a_\mu(\mathbf{k})+i k_\mu k\cdot \widetilde{\zeta}(k).
\label{god}
\end{eqnarray}
Thus if we consider tree diagrams with $N$ electrons and photons emitted with varying momenta, when an electron line changes momentum from $p$ to $p'$, the vertex involved will carry the factor
$\gamma\cdot (p-p') (p-p')\cdot \widetilde{\zeta}(p-p')$. Since varying electron momenta will occur, the $\zeta$ will not factor out and  the response of {\it in} state to $e^{i Q(\zeta)}$ will be non-trivial.

But this eigenvalue cannot change as $ Q(\zeta)$ is superselected. That should give identities for scattering amplitudes involving photon momenta $k$, $k', \cdots$.
Such a calculation is beyond the scope of this paper.

\section {The BMS Group}
The Bondi-Metzner-Sachs group was introduced during the study of classical gravitational radiation \cite{bms,bms1}.
It acts on null infinity $\mathcal{J}^+$. We explain its action to the extent we require. See also \cite{Nilsson}.

The four-dimensional conformal group $SO(4,2)$ does not act on Minkowski space $M_4$. It acts only on the Dirac-Weyl compactification $\overline{M}_4$ of $M_4$.
Consider the six dimensional space $M_{4,2}$ with topology $\mathbb{R}^6$, coordinates $(\xi_0,\xi_1,\dots, \xi_4,\xi_5)=(\xi_\mu,\xi_4,\xi_5)$ and metric $\langle\xi\rangle=\xi^2_0-=\sum_{i=1}^4\xi^2_i$. We can write this metric as
\begin{equation}
\xi^\mu\xi_\mu-(\xi_4+\xi_5)(\xi_4-\xi_5).
\end{equation}
The null cone in $M_{4,2}$ is 
\begin{equation}
V=\left\{\xi: \xi^\mu \xi_\mu- (\xi_4+\xi_5)(\xi_4-\xi_5)=0\right\}
\end{equation}
Let us consider $\mathbb{V P}{}$, the projective space associated with $V$:
\begin{equation}
\mathbb{VP}{}=\left\{\langle\xi\rangle: \langle\xi\rangle=\langle\lambda \xi\rangle\ ; \xi\in V, \lambda\neq 0\right\} \ 
\end{equation}
$\mathbb{VP}{}$ is a four-dimensional space.

Let 
\begin{equation}
\xi_4=\frac{\gamma}{2}\left(1+\frac{\xi\circ\xi}{\gamma^2}\right), \quad \xi_5=\frac{\gamma}{2}\left(1-\frac{\xi \circ\xi}{\gamma^2}\right),\quad \xi\circ\xi:=\xi^\mu\xi_\mu.
\label{bmsc}
\end{equation}
Then $\xi\in \mathbb{VP}.$

If $\gamma\neq 0$
\begin{equation}
\langle\xi\rangle=\langle {\textstyle \frac{\xi}{\gamma}}, {\textstyle \frac12} (1+{\textstyle \frac{\xi}{\gamma}}\circ {\textstyle\frac{\xi}{\gamma}}),{\textstyle \frac12}(1-{\textstyle\frac{\xi}{\gamma}}\circ{\textstyle \frac{\xi}{\gamma}})\rangle ,
\label{bms0}
\end{equation}
where on L.H.S., $\xi$ is ${\xi_\mu}$.

Setting 
$$ x^\mu=\frac{\xi^\mu}{\gamma},\qquad \gamma\neq 0,$$
we see that the {\it interior } of $\mathbb{VP}{}$ with $\gamma\neq 0$ is the Minkowski space:
\begin{equation}
\gamma_0\neq 0: \langle\xi\rangle=\langle  x, {\textstyle \frac12} (1+x\cdot x),{\textstyle \frac12}(1-x\cdot x)\rangle.
\label{bms1}
\end{equation}
But if $\gamma=0$, $\xi_4+\xi_5=0$ and 
\be \xi=(\xi^\mu,\xi_4,-\xi_4), \qquad \xi^\mu\xi_\mu=0. 
\ee
Thus if $\xi_4\neq0$, $\langle\xi\rangle$ spans a light cone for each sign of $\xi^0/\xi_4$
\begin{equation}
\xi_4\neq 0: \langle\xi\rangle=\langle  \xi^\mu/\xi_4,1,-1\rangle.
\label{bms2}
\end{equation}

Let us call this space as $\mathcal{J}$. We can regard $\xi_4=0$ as its tip.It has the topology of $S^2$. It is obtained by blowing up the origin $\xi^\mu=0$.

The BMS group acts on $\mathcal{J}$. For convenience let us set $\xi^\mu/\xi_4=N^\mu$ and distinguish
$N^0\gtrless0$,
\begin{equation}
\mathcal{J}^\pm=\left\{ N, N\circ N=0; N^0 \gtrless 0,\right\}.
\label{bm3}
\end{equation}
The BMS group acts on $\mathcal{J}^\pm$. We focus on $\mathcal{J}^+$. 
With $N^0>$, we can write 
\be N=(N^0,N^0\mathbf{N}), \quad \mathbf{N}\cdot \mathbf{N}=1.\ee
Thus,
\begin{equation}
\mathcal{J}^\pm=\mathbb{R}\times S^2
\label{bm4}
\end{equation}
with coordinates 
\begin{equation}
(N^0,\mathbf{N}).
\label{bm5}
\end{equation}
The BMS group consists of a pair $(\alpha, \Lambda)$, where $\alpha$ is a real function on $S^2$,
\begin{equation}
\alpha:S^2\rightarrow \mathbb{R}.
\label{bm6}
\end{equation}
and $\Lambda$ is a Lorentz transformation
\begin{equation}
\Lambda\in \mathcal{L}_+^\uparrow.
\label{bm7}
\end{equation}
The action of $(\alpha,\Lambda)$ on $(N^0,\mathbf{N})$ is 
\begin{equation}
(\alpha,\Lambda)(N^0,\mathbf{N})=(N^0+\alpha( \Lambda\circ \mathbf{N}), \Lambda\circ \mathbf{N}),
\label{bm8}
\end{equation}
where $\Lambda\circ \mathbf{N}$ denotes the action of $\mathcal{L}_+^\uparrow$  on $S^2$ as
conformal transformation.

Let $\alpha\to \Lambda^\ast \alpha$ be the usual pull-back action of $\mathcal{L}_+^\uparrow$ on $\alpha$

\begin{equation}
\Lambda^\ast \alpha( \mathbf{N})=\alpha(\Lambda\circ  \mathbf{N}).
\label{bm9}
\end{equation}

Then (\ref{bm8}) shows that the BMS group is the semi-direct product of $\mathcal{L}_+^\uparrow$ with {\it supertranslations} $\alpha$. The composition law for the latter is addition of functions so that it is abelian. We find 
\begin{equation}
(\alpha_1, \Lambda_1)(\alpha_2, \Lambda_2)=(\alpha_1+ \Lambda_1^\ast \alpha_2, \Lambda_1\Lambda_2 ).
\label{bm10}
\end{equation}
The subalgebra where $\alpha$ has just angular momenta $0$ and $1$,
\begin{equation}
\alpha( \mathbf{N})=a^0+  \mathbf{a}\cdot \mathbf{N};\qquad a^\mu \in  \mathbf{R}^4.
\label{bm11}
\end{equation}
gives the Poincar\'e group.

\vspace{12pt}
\parindent=18pt{$\bullet\ $} {\it Away from $\mathcal{J}^+$}
\vspace{12pt}

In quantum theory, it is important to realize the BMS group as operators on the quantum Hilbert space. In our case, the latter carries a non-trivial representation of QED because of infrared effects.

We can approach the problem by extending the action of $\mathcal{J}^+$ to all of $\overline{M}_4$.  With that in hand, we can transform test functions and perhaps find operators to implement these transformations.

But the BMS group acts only on  $\mathcal{J}^+$. There are many ways to extend its action to ~$\overline{M}_4$.

We now suggest that the BMS does act on $\gc/\gci$, but not on $\gc$ or $\gci$ separately. The reason is as follows. Let us characterise elements of $\gci$ as in (\ref{t4}),
but dropping (\ref{t5}). It is convenient to do so, and the results are unaffected.

The action of the BMS on the leading asymptotic terms (\ref{t6}) is fixed. Hence if $(\alpha,\Lambda)$, $(\alpha,\Lambda)'$ are two such actions of this group on $\widetilde{\zeta}_\mu$ which however coincide on the asymptotic terms, then
\begin{equation}
(\alpha,\Lambda)\ k\cdot \widetilde{\zeta}^(\mathbf{k})- (\alpha,\Lambda)'\ k\cdot \widetilde{\zeta}^(\mathbf{k})
\xrightarrow[k\to 0]{}0
\label{bms66}
\end{equation}
Thus, 
\begin{equation}
Q((\alpha,\Lambda)\ {\zeta})- Q((\alpha,\Lambda)'\  {\zeta})
= G((\alpha,\Lambda)\ {\zeta}-(\alpha,\Lambda)'\ {\zeta}')
\label{bms666}
\end{equation}
is a generator of $\gci$, and vanishes on quantum states.

All possible extensions of $(\alpha,\Lambda)$  from boundary to bulk act in the same manner on $\gc/\gci$

Thus

 \parindent=18pt{$\bullet\ $}{\it BMS acts on the superselection Algebra}

We now give an example of an extension of the  BMS group  action from $\mathcal{J}^+$ to $\overline{M}_4$.

Consider (\ref{comm}) which give the gauge transformation of $Q(\zeta)$ and hence defines $\gc$- Substituting
\begin{equation}
u=x_0-r\ ,
\label{bms7}
\end{equation}
we get 
\begin{equation}
e^{\pm i k\cdot x}= e^{\pm i \left[k^0(u+r)-k^0\, r\, \hat{\mathbf{k}}\cdot\hat{\mathbf{x}}\right]}.
\label{bms8}
\end{equation}
For a generic supertranslation,
$$r\to r+\frac{\alpha(\hat{r})}{k^0}$$
so that it survives the $k_0\to0$   limit in ${\mathbf k}\cdot {\mathbf x}$. We can replace  $k_0$ by any smooth function $\beta(k_0)$
 which  goes like $k^0$ as $k_0\to0$ and say vanishes fast as $k_0$ increases.That transforms $\partial_\mu\Lambda$ to a new function $(\alpha,\Lambda)\partial_\mu\Lambda=\partial_\mu\Lambda'$ defined for all $\Lambda$ as we require.
 
 There are no divergences in (\ref{comm}) for $n\leq2$.
 
 \vspace{12pt}
\parindent=18pt{$\bullet\ $} {\it The BMS group is spontaneously broken}
\vspace{12pt}

The transformation of  $\partial_\mu\Lambda$ to $\partial_\mu\Lambda'$ by supertranslations or boosts is generically non-trivial and changes it at infinity. Thus this action 
 is non-trivial on $\gc/\gci$ and changes the superselection sector. The exception is the rotation subgroup which acts trivially $Q(\zeta)$ because $d\mu({k}) $ is a rotationally invariant measure.
 
 Hence, the BMS group is spontaneously broken to its rotation subgroup.

\section{Conclusions} { }
We have shown that equations of motion can be considered as constrains in field theory.
This interpretation allows us to define a covariant version of Gauss law.
Using the Peierls' formulation of quantization, we analysed the physics 
effects of the infrared behaviour of QED.
In particular we have shown that the  infrared dressed one-particle 
states induce a spontaneous symmetry breaking 
of some space-time symmetries like Lorentz transformations because they change the charged superselection sector. 

The same analysis affects the role of other
asymptotic {\it symmetry} groups such as BMS which act on the {\it boundaries} of spacetime.
However, because gauge invariance under local gauge transformation are preserved by the covariant
Gauss law constraints, Ward identities  of the S-matrix under $\mathcal{G}_0^\infty$ still hold which permits us to  generalize the standard results to deduce to charge conservation and low energy theorems.

A crucial observation is that the proof of such results does not requires Lorentz invariance.

\subsection*{Acknowledgements}
This article is based upon work from COST Action MP1405 QSPACE, supported by COST (European Cooperation in Science and Technology). M.~Asorey  work has been partially supported by the Spanish MINECO/FEDER grant FPA2015-65745-P
and DGA-FSE grant 2015-E24/2.  F.L.\ is supported by INFN, I.S.'s GEOSYMQFT and received partial support by CUR Generalitat de Catalunya under projects FPA2013-46570 and 2014~SGR~104, MDM-2014-0369 of ICCUB (Unidad de Excelencia `Maria de Maeztu').

\end{document}